# A Non-Parametric Test of Risk Aversion

Jacob K. Goeree and Bernardo García-Pola[1]

August 7, 2023

**Abstract**

In economics, risk aversion is modeled via a concave Bernoulli utility within the expected-utility paradigm. We propose a simple test of expected utility and concavity. We find little support for either: only 30% of the choices are consistent with a concave utility, only two out of 72 subjects are consistent with expected utility, and only one of them fits the economic model of risk aversion. Our findings contrast with the preponderance of seemingly "risk-averse" choices that have been elicited using the popular multiple-price list methodology, a result we replicate in this paper. We demonstrate that this methodology is unfit to measure risk aversion, and that the high prevalence of risk aversion it produces is due to parametric misspecification.

[1]Goeree: AGORA Center for Market Design, UNSW, Sydney, Australia. García-Pola: Department of Economics, Universidad Pública de Navarra, Pamplona, Spain. We gratefully acknowledge funding from the Australian Research Council (DP190103888, DP220102893). We thank Peter Wakker and Brett Williams for helpful comments, Filip Fidanoski and Andreas Ortmann for an overview of the many risk-elicitation methods, and Charles Holt for sharing instructions for the Holt–Laury task.

# 1. Introduction

Risk attitudes play an important role in economic and financial decisionmaking. Risk aversion gives rise to demand for insurance, impacts consumption and saving levels, shapes moral hazard and the design of contracts, and affects bidding in pay-your-bid auctions. In finance, risk attitudes are key when evaluating the trade-off between an asset's risk and return and when composing investment portfolios. Because of the ubiquitous role of risk in decisionmaking, the proper measurement of risk preferences is important for economic analysis, policy, and financial advice.

Financial institutions typically use surveys and self reports to assess clients' risk preferences.[1] The downside is that the data so obtained are qualitative in nature and do not lend themselves to extrapolation beyond the survey's hypothesized scenarios. Incentivized experiments offer a viable alternative for the quantitative measurement of risk preferences. Holt and Laury (2002) proposed a simple task based on the multiple-price list methodology that has a long tradition in development economics and psychology.[2] Their seminal contribution led to a surge of interest in measuring risk attitudes in the laboratory. In the past two decades, a wide variety of risk elicitation methods have been proposed by experimental economists.[3]

However, results from these elicitation methods do not only significantly differ from those of surveys, e.g. Anderson and Mellor (2009), they also vary considerably across the different methods and sometimes contradict each other. There is no consensus why the results vary across elicitation methods nor which elicitation method is most accurate.[4] This lack of consistency, dubbed the *risk elicitation puzzle* by

---

[1]Many countries require financial institutions to gauge investors' risk attitudes. For instance, the European Securities and Markets Authority dictates that "firms should specify the general attitude that target clients should have in relation to the risks of investment" (Guidelines on MiFID II, 2018).

[2]See, e.g., Binswanger (1980) and Tversky and Kahneman (1982, p.305-306). The brief description of the multiple-price list methodology in the latter underlines it was considered standard at the time. We thank Peter Wakker (private communication) for pointing out that multiple-price lists have been used in decision theory as long as the field exists.

[3]See, e.g., Holt and Laury (2002, 2005); Eckel and Grossman (2002, 2008); Choi et al. (2007); Crosetto and Filippin (2013); Dohmen et al. (2010); Abdellaoui et al. (2011). This list is far from complete. Fidanoski and Ortmann (private communication) report there are over a 100 different risk-elicitation methods.

[4]See, e.g., Charness et al. (2013); Crosetto and Filippin (2015); Pedroni et al. (2017); Holzmeister and Stefan (2021). Friedman et al. (2022) show that task attributes, e.g. text versus visualization and continuous versus discrete choices, explain some of the (lack of) correlation across various tasks.

Pedroni et al. (2017), undermines the external validity of laboratory risk-elicitation methods (Smith, 1989; Friedman et al., 2014).

Despite the large number of risk-elicitation methods that have been proposed, none of them directly tests the two key assumptions that underlie the economic model of risk aversion. Expected utility is linear in a lottery's probabilities and the lottery's prizes are evaluated using a concave Bernoulli utility. We introduce a simple task that entails comparing a baseline lottery to two related lotteries. This task allows us to test for concavity of the Bernoulli utility without making parametric assumptions. Moreover, by varying the probabilities of the baseline lottery it allows us to test the linearity assumption. Our experimental results refute both concavity and linearity.

Our results contrast with a large body of experimental work that finds that the vast majority of choices are "risk averse" when using the task proposed by Holt and Laury (2002). This finding, however, hinges on parametric assumptions about the Bernoulli utility, e.g. that it exhibits constant relative risk. We demonstrate that the Holt–Laury task is unfit to test for risk aversion, and that the high prevalence of "risk averse" choices it produces is due to parametric misspecification.

The next section proposes a simple non-parametric test of expected utility and risk aversion. Section 3 reports an experiment based on this test. Section 4 discusses the multiple-price list methodology. Section 5 concludes. Appendix A contains the proof of Theorem 1 and Appendix B contains the instructions for the experiment.

## 2. Mean-Preserving Spreads

A classic result due to Rothschild and Stiglitz (1970) is that *any* risk averse individual prefers a lottery to a mean-preserving spread of itself. Surprisingly, this result has hitherto not been systematically exploited in the large literature on risk elicitation.

Consider lotteries over four prizes $\pi = (\$1, \$16, \$21, \$38.5)$ with utilities $u(1)$, $u(16)$, $u(21)$, and $u(38.5)$.[5] We can subtract a constant from these utilities or multiply them by a positive constant and the result would describe the same preferences. So let us normalize the utilities as 0, $u_1$, $u_2$, and 1 where $0 \leq u_1 \leq u_2 \leq 1$.[6] This utility-possibility set corresponds to the triangular area above the 45-degree line in Figure 1. The Bernoulli utility is concave if and only if its slope is non-increasing, i.e.

---

[5]These prizes were chosen to match those of the Holt–Laury task, see Section 4 and Appendix B.
[6]This follows by subtracting $u(1)$ from all utilities and dividing the resulting utilities by $u(38.5) - u(1)$ so that $u_1 = (u(16) - u(1))/(u(38.5) - u(1))$ and $u_2 = (u(21) - u(1))/(u(38.5) - u(1))$.

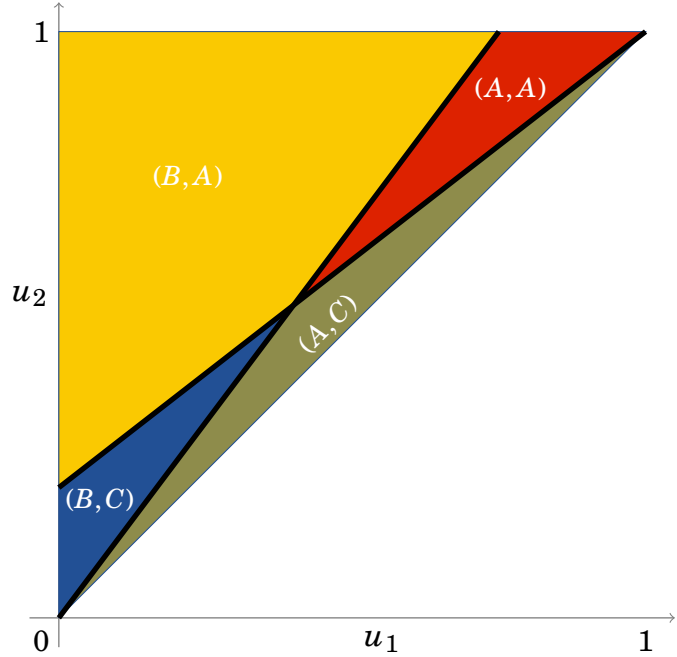

**Figure 1:** The colored areas represent possible utility pairs $(u_1, u_2)$ for each of the four outcomes when an individual chooses between lotteries $(\mathscr{L}_A, \mathscr{L}_B)$ and then between lotteries $(\mathscr{L}_A, \mathscr{L}_C)$. The red area corresponds to concave Bernoulli utilities.

$$\frac{u_1}{15} \geq \frac{u_2 - u_1}{5} \geq \frac{1 - u_2}{17.5} \tag{1}$$

where the numerators are differences in utilities and the denominators are differences in prizes. The black lines in Figure 1 show utility pairs $(u_1, u_2)$ for which these inequalities hold with equality. The red area shows utility pairs such that both inequalities in (1) hold and the blue area shows utility pairs for which both inequalities are reversed. The yellow (green) area shows utility pairs for which only the first (second) inequality of (1) is reversed.

Suppose lottery $\mathscr{L}_A$ offers the prizes $\pi$ with probabilities $p = (p_1, p_2, p_3, p_4)$. One possible mean-preserving spread of $\mathscr{L}_A$ results from moving all mass of the second prize to the first and third prizes. This yields lottery $\mathscr{L}_B$ with probabilities $p = (p_1 + \frac{1}{4}p_2, 0, p_3 + \frac{3}{4}p_2, p_4)$ over the same prizes $\pi$. The reason that only one-quarter of the mass is moved downward and three-quarters are moved upward is to keep the mean the same. A second mean-preserving spread of $\mathscr{L}_A$ results from moving all mass of the third prize to the second and fourth prizes, which results in lottery $\mathscr{L}_C$ with probabilities $p = (p_1, p_2 + \frac{7}{9}p_3, 0, p_4 + \frac{2}{9}p_3)$.[7]

Consider the task of choosing between lotteries $(\mathscr{L}_A, \mathscr{L}_B)$ and between lotteries $(\mathscr{L}_A, \mathscr{L}_C)$. If $\mathscr{L}_A$ is preferred over $\mathscr{L}_B$ then $u_1 \geq \frac{3}{4}u_2$, which is equivalent to the first inequality in (1). If $\mathscr{L}_A$ is preferred over $\mathscr{L}_C$ then $u_2 \geq \frac{7}{9}u_1 + \frac{2}{9}$, which is equivalent to the second inequality in (1). So, if $\mathscr{L}_A$ is chosen twice, the utility pair $(u_1, u_2)$ belongs to the red area and the Bernoulli utility is concave. If $\mathscr{L}_A$ is not chosen at all then the utility pair belongs to the blue area and the Bernoulli utility is convex. If $\mathscr{L}_B$ and $\mathscr{L}_A$ are chosen then the utility pair belongs to the yellow area, which reflects a Bernoulli utility that is steepest for intermediate prizes. Finally, if $\mathscr{L}_A$ and $\mathscr{L}_C$ are chosen then the utility pair belongs to the green area, which reflects a Bernoulli utility that is least steep for intermediate prizes. Bernoulli utilities belonging to the green and yellow areas are neither convex nor concave.

The analysis of the previous paragraph applies regardless of the probabilities that define lottery $\mathscr{L}_A$. This is a consequence of the linearity assumption that underlies expected utility. We can test for both linearity and concavity by letting individuals choose between $(\mathscr{L}_A, \mathscr{L}_B)$ and $(\mathscr{L}_A, \mathscr{L}_C)$ for various values of $p = (p_1, p_2, p_3, p_4)$. The next theorem generalizes to lotteries over an arbitrary number of prize values.

**Theorem 1** *Consider any lottery $\mathscr{L}$ over $K > 1$ ascending prizes $\pi = (\pi_1, \ldots, \pi_K)$ that occur with probabilities $p = (p_1, \ldots, p_K)$ where $p_k$ is strictly positive for $1 < k < K$. Let $\mathscr{L}_k$ for $1 < k < K$ denote the lotteries in which all probability mass of prize $k$ is moved to the prizes $k-1$ and $k+1$ in a mean-preserving manner. An individual is risk averse if and only if lottery $\mathscr{L}$ is preferred to lottery $\mathscr{L}_k$ for all $1 < k < K$.*

Theorem 1 offers a simple non-parametric test of risk aversion that involves $K - 2$ comparisons between two lotteries. Combined they yield necessary and sufficient conditions for concavity of the Bernoulli utility (see Appendix A).[8] Varying the probabilities that define the lottery $\mathscr{L}$ in Theorem 1 yields a simple test of expected utility.

---

[7]Note that we cannot move mass of the extreme prizes $1 and $38.5 to any of the other prizes while keeping the mean the same.

[8]The few existing non-parametric tests of risk aversion typically exploit an implication of concavity, rather than testing for concavity directly. Baillon and L'Haridon (2021) propose a non-parametric version of the well-known Arrow–Pratt measure. Johnson et al. (2021) measure the utility function using the adaptive trade-off method proposed by Wakker and Deneffe (1996). L'Haridon and Vieider (2019) elicit risk preferences and prospect theory parameters using certainty equivalents. Our task, which entails two choices between two lotteries, is simpler and avoids the certainty effect.

# 3. Experiment

Section 3.1 describes the experimental procedures and protocol. The two main hypotheses can be found in Section 3.2 and Section 3.3 discusses the results.

## 3.1. Experimental Design and Protocol

We conducted 7 sessions and recruited a total of 72 subjects using ORSEE (Greiner, 2015). The sessions were conducted online via Zoom, and the experimental tasks were administered using z-Tree, see Fischbacher (2007), and z-Tree Unleashed, see Duch et al. (2020).

The sessions were divided into two parts. In the first part, instructions for the Holt–Laury task were shown using a PowerPoint presentation that was read aloud, see Appendix B. Subjects' screens showed a list of ten scenarios asking them to choose between a "safe" and "risky" lottery. Once participants finished the first part, we read the instructions for part two out loud. In this part, subjects went through six different screens, each of them displaying different three lotteries $(\mathscr{L}_A, \mathscr{L}_B, \mathscr{L}_C)$. For each screen, subjects had to choose one lottery from the pair $(\mathscr{L}_A, \mathscr{L}_B)$ and one lottery from the pair $(\mathscr{L}_A, \mathscr{L}_C)$, resulting in a total of twelve decisions in the second part. The position of the lotteries, decisions, and buttons were randomized on the screens. This randomization was done independently for each subject.

Participants had the opportunity to ask questions during both the instruction phase and the actual experiment. At the end of the experiment, we determined subjects' earnings and paid them in private. Each subject was paid for one randomly chosen scenario from part one and one randomly chosen decision from part two. On average, participants earned approximately AU$49, including the show-up fee.

Table 1 shows the lotteries used in the second part of the experiment. In each of the six cases, the prize values were $\pi = (\$1, \$16, \$21, \$38.5)$. The second column shows the probabilities that define lottery $\mathscr{L}_A$. The third and fourth columns show the lotteries $\mathscr{L}_B$ and $\mathscr{L}_C$ that follow by moving all mass of the second and third prize respectively. For each of the six cases in Table 1, subjects chose one lottery from the pair $(\mathscr{L}_A, \mathscr{L}_B)$ and one lottery from the pair $(\mathscr{L}_A, \mathscr{L}_C)$.

| Case | $\mathcal{L}_A$ | $\mathcal{L}_B$ | $\mathcal{L}_C$ |
|------|-----------------|-----------------|-----------------|
| C1 | (21, 16, 63, 0) | (25, 0, 75, 0) | (21, 65, 0, 14) |
| C2 | (27, 64, 9, 0) | (43, 0, 57, 0) | (27, 71, 0, 2) |
| C3 | (57, 16, 27, 0) | (61, 0, 39, 0) | (57, 37, 0, 6) |
| C4 | (0, 16, 63, 21) | (4, 0, 75, 21) | (0, 65, 0, 35) |
| C5 | (0, 64, 27, 9) | (16, 0, 75, 9) | (0, 85, 0, 15) |
| C6 | (0, 16, 27, 57) | (4, 0, 39, 57) | (0, 37, 0, 63) |

**Table 1:** Probabilities (in percentages) of the lotteries used in the second part of the experiment.

### 3.2. Hypotheses

The four possible outcomes for each of the six cases in Table 1 are $(A,A)$, $(A,C)$, $(B,A)$, or $(B,C)$, corresponding to the colored areas of Figure 1. The outcome is referred to as the subject's choice.

**Hypothesis 1 (Expected Utility)** *A subject's choice is the same in all six cases.*

In other words, a subject's choice may fall into any of the four colored regions of Figure 1 as long as it consistently falls into the same region across all six cases.

The next hypothesis reflects the finding of numerous risk-elicitation experiments that the bulk of subjects' choices are risk averse. For our task, this means that subjects' choices fall into the red area in Figure 1 since $(A,A)$ is the only choice compatible with a concave Bernoulli utility.

**Hypothesis 2 (Risk Aversion)** *The vast majority of subjects' choices is $(A,A)$.*

Finally, the economic model of risk aversion applies when both hypotheses hold.

### 3.3. Results

Table 2 shows the results for the second part of the experiment. The bars display the proportion of subjects that chose $(A,A)$, $(B,A)$, $(A,C)$, or $(B,C)$ for each of the six cases of Table 1. Before testing the hypotheses of the previous section we rule out that behavior is random, either because of risk neutrality or confusion.

**Result 1** *Choices in the second part of the experiment are not random.*

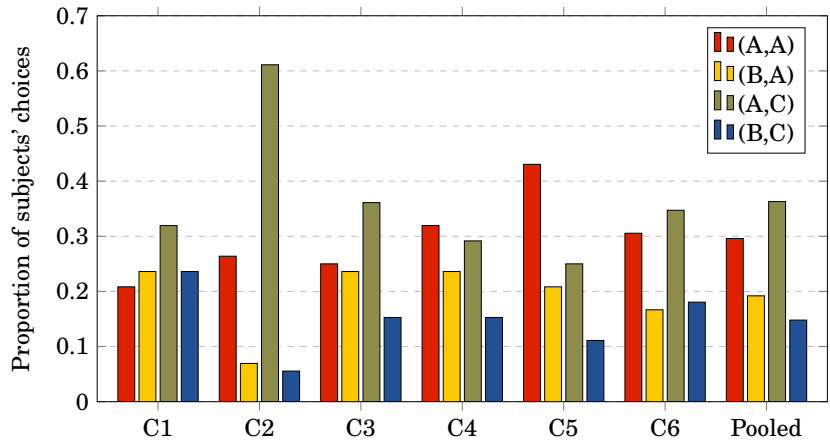

**Figure 2:** Proportion of subjects' choices for the six cases of Table 1.

**Support:** The pooled percentages across the six cases are 29.6% for $(A,A)$, 19.2% for $(B,A)$, 36.3% for $(A,C)$, and 14.8% for $(B,C)$, see the rightmost bars in Figure 2. A Pearson $\chi^2$ test reveals that these percentages are significantly different ($p < 0.0001$) from those expected under random behavior (25% for each choice). ∎

**Result 2** *Hypothesis 1 is rejected.*

**Support:** Subject's choices differ across the six cases of Table 1 at both the individual and population levels. Out of 72 subjects, only two make the same choice across all six cases, and only a minority of subjects (31) make the same choice in more than half of the cases. Even if we ignore individual consistency and consider the distributions of choices, a Pearson's $\chi^2$ test reveals they differ across the six cases ($p = 0.0004$). ∎

**Result 3** *Hypothesis 2 is rejected.*

**Support:** Pooling across the six cases of Table 1, the percentage of $(A,A)$ choices that belong to the red area is only 29.6%. A majority of the choices belong to the yellow and green areas, i.e. they are neither risk averse nor risk loving. Among the sixteen most consistent subjects (that make the same choice in five or six cases) the percentage of choices that belong to the red area is even lower, 25.0%, while 68.8% of the choices belong to the yellow and green areas. ∎

Result 3 contrasts with the preponderance of seemingly "risk averse" choices that are typically found using the multiple-price list methodology. We next discuss the reason for this discrepancy.

## 4. Multiple-Price Lists

Holt and Laury (2002) measure risk aversion using a multiple-price list. Subjects choose between a "safe" lottery with prizes of $16 and $21 and a "risky" lottery with prizes of $1 and $38.5. They face a total of ten "safe" versus "risky" lottery comparisons that result by varying the probability of the best outcome ($21 for the safe lottery and $38.5 for the risky lottery) from $p = 0.1$ to $p = 1.0$, see Appendix B.[9] Multiple-price lists are popular as they are easy to use and interpret. Subjects typically choose the safe lottery when $p$ is below some threshold and then switch to the risky lottery. The threshold is then used to parametrically identify risk aversion.

The Holt–Laury task has become the "gold standard" for risk elicitation. However, we will demonstrate that inferences based on this task reflect parametric assumptions rather than risk aversion per se. To see this, let us use the same normalization of the utilities for the four prizes as in Section 2. If an individual makes $s$ safe choices and then switches, the utilities satisfy

$$1 - \frac{10-s}{s} u_1 \leq u_2 \leq 1 - \frac{9-s}{s+1} u_1 \qquad (2)$$

where the first inequality follows since the safe lottery was chosen when $p = \frac{s}{10}$ and the second inequality follows since the risky lottery was chosen when $p = \frac{s+1}{10}$. The inequalities in (2) define the colored triangles in Figure 3. In this figure, the dark blue triangle on the left corresponds to zero safe choices and the dark red triangle at the top corresponds to nine safe choices. The white number in each triangle indicates how many subjects in our experiment (out of 72) made the number of safe choices corresponding to that triangle. In line with previous studies, we find that a vast majority (69.4%) of subjects make five or more safe choices.

Holt and Laury (2002) conclude these subjects are risk averse by assuming their utilities belong to the black curve, which shows all possible constant relative risk utilities (i.e. for any value of the constant relative risk parameter $r$). For instance, for the 12 subjects that made six safe choices the inferred parameter is in the range $r \in [0.37, 0.64]$. The lower number of this range corresponds to the value of $r$ for which the black curve enters the triangle from the left and the higher number corresponds

---

[9]One difference with the Holt and Laury (2002) study is that in our experiment, subjects could switch between the safe and risky lotteries only once.

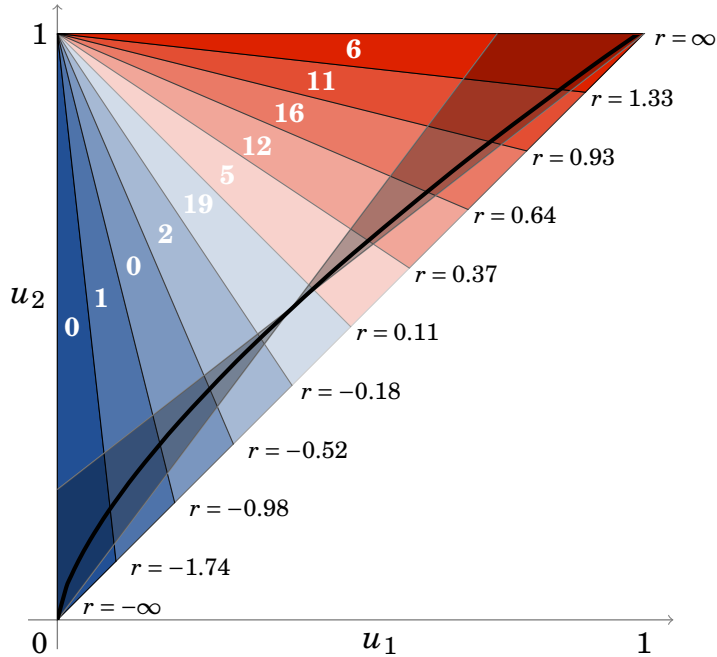

**Figure 3:** The colored triangles show the possible utilities based on the number of "safe" choices ranging from zero (dark blue triangle) to nine (dark red triangle). The white number in each triangle indicates how many subjects (out of 72) make the number of safe choices corresponding to that triangle. The upper shaded area corresponds to concave utilities and the lower shaded area to convex utilities, cf. the red and blue areas in Figure 1. The black curve corresponds to constant relative risk utilities $u(x) = x^{1-r}/(1-r)$ for $r \in \mathbb{R}$. The lower corners of each triangle indicate the range of $r$ values for which that triangle applies (assuming constant relative risk).

to the value of $r$ for which the black curve exits the triangle on the right.

However, any of the triangles in Figure 3 overlaps the green and yellow areas in Figure 1. This means that Bernoulli utilities belonging to either of these areas can explain *any* number of safe choices in the Holt–Laury task. Economic theory does not preclude such utilities. For instance, subjects may value similar prizes like $16 and $21 similarly, and differently from more extreme prizes like $1 and $38.5. This corresponds to the green area, i.e. a utility that is neither concave nor convex as it is least steep in the middle. Such a utility implies risk seeking if there is a chance of the best prize ($38.5) and risk aversion if there is a chance of the worst prize ($1). Alternatively, subjects may be risk seeking over the "loss" of getting $1 and risk averse over the "gain" of getting $38.5, which corresponds to a utility in the yellow area.

While utilities in the green and yellow areas are plausible and perfectly consis-tent with expected utility theory, they are ruled out by the assumption of constant relative risk. Under this assumption, each of the triangles in Figure 3 is projected onto the black curve, despite the triangle containing Bernoulli utilities that are neither concave nor convex. Indeed, the *majority* of utilities belong to either the green or yellow area, see Result 3. By imposing constant relative risk, utilities from these areas are misspecified. This misspecification raises the number of "risk averse" choices from 30% to almost 70%, thus exaggerating the prevalence of risk aversion.

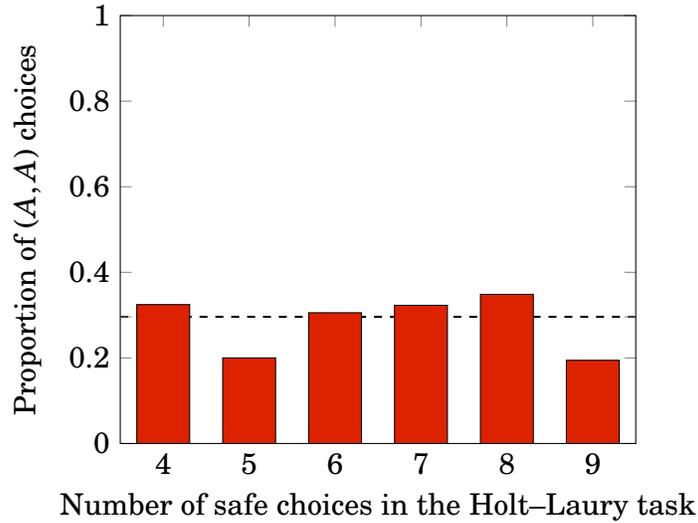

**Figure 4:** The bars show the proportion of $(A,A)$ choices by the number of safe choices in the Holt–Laury task and the dashed line shows the overall average of $(A,A)$ choices (29.6%).

The next result reflects that *all* triangles in Figure 3 overlap with the yellow and green areas in Figure 1.

**Result 4** *The number of safe choices in the Holt–Laury task is not correlated with concavity of the Bernoulli utility.*

**Support:** Figure 4 shows there is no correlation between the number of safe choices in the Holt–Laury task and the choices that correspond to a concave utility. A Pearson $\chi^2$ test confirms that the distribution in Figure 4 is not significantly different from one that is constant and equal to the overall average of 29.6% ($p = 0.64$). (For those classified as "risk averse" under the Holt–Laury task, i.e. those with five or more safe choices, the average percentage of $(A,A)$ choices is 29.7%.) ∎

Finally, under the Holt–Laury task, little can be gained by considering alternative models such as constant absolute risk or the "power-expo" model used by Holt and Laury (2002). These alternatives may provide a better fit of the aggregate choice data, but are also prone to misspecification as they simply generate a different curve in the shaded triangles of Figure 3. Imposing any of these alternative models is tantamount to projecting the non-convex and non-concave utilities in the green and yellow areas of Figure 1 onto this curve.

To summarize, the Holt–Laury task is unfit to measure risk aversion of any kind since the inequalities it generates result in the blue and red triangles of Figure 3, rather than the blue and red areas of Figure 1.

## 5. Discussion

Following Holt and Laury (2002), and continuing a long tradition in psychology, multiple-price lists have been employed in hundreds of risk-elicitation experiments. They may well be the most commonly-used methodology in experimental economics today. One reason for their frequent use is that risk-elicitation tasks are routinely appended to experiments in which risk aversion is thought to play a role. Another is that multiple-price lists are easy to implement and (seemingly) easy to interpret. Assuming constant relative risk, the number of safe choices in the Holt–Laury task pins down a lower and upper bound for the risk aversion parameter, see Figure 3.[10]

One methodological contribution of this paper is to elucidate the role of this assumption in the finding that the vast majority of choices are risk averse. The four-prize setup of the Holt–Laury task is simple enough to graph the risk-loving and risk-averse Bernoulli utilities as subsets of the set of all non-decreasing utilities, see the blue and red areas in Figure 1. There is no a priori reason that subjects' utilities belong to these areas. Expected utility theory allows for Bernoulli functions that are neither convex nor concave, as indicated by the green and yellow areas in Figure 1.

---

[10]The Holt–Laury task is not the only one that relies on parametric assumptions. There are now over a hundred different elicitation methods and many of them invoke the assumption of constant relative risk, see e.g. Andersen et al. (2008); Eckel and Grossman (2008); Anderson and Mellor (2009); Crosetto and Filippin (2013); Holzmeister and Stefan (2021). As do the survey papers that compare different elicitation methods, see e.g. Charness et al. (2013); Csermely and Rabas (2016); Pedroni et al. (2017); Eckel (2019); Holzmeister and Stefan (2021).

Imposing constant relative risk results in misspecification. This follows from a geometric (rather than econometric) argument. The Holt–Laury task generates the triangles in Figure 3. The red triangles are assumed to reflect risk aversion even though they overlap with the green and yellow areas in Figure 1 that contain utilities that are neither convex nor concave. Mapping these areas onto the constant relative risk curve in Figure 3 exaggerates the case for risk aversion. In other words, the high prevalence of "risk averse" choices in the Holt–Laury task reflects the *assumption* of constant relative risk, but does not constitute proof of risk aversion per se.

A second methodological contribution is that we introduce a simple task that generates the correct inequalities to test for concavity, i.e. whether utilities belong to the red area in Figure 1. It is based on Rothschild and Stiglitz's (1970) classic result that *any* risk averse individual prefers a lottery to a mean-preserving spread of itself. By varying the probabilities of the lottery we can further test the linearity assumption that underlies expected-utility theory.

A third contribution is empirical. Our task leads to non-random behavior that shows little evidence of risk aversion, see Results 1 and 3. Only 30% of the choices belong to the red area of Figure 1, while a majority of the choices belong to the yellow and green areas. In line with many previous experiments, we replicate that a vast majority of subjects make five or more safe choices in the Holt–Laury task. However, these "risk averse" choices mostly reflect utilities that are neither convex nor concave. There is no correlation between the number of safe choices in the Holt–Laury task and concavity of the utility, see Result 4. Finally, we find little evidence for the linearity assumption that underlies the expected-utility hypothesis, see Result 2.

# A. Proof of Theorem 1

The "only if" part of Theorem 1 follows from Rothschild and Stiglitz's (1970) classic result that a risk averse individual prefers any lottery to a mean-preserving spread of itself. For the "if" part we need to show that the inequalities generated by the choices between $(\mathscr{L}, \mathscr{L}_k)$ for $1 < k < K$ are sufficient to establish concavity of the Bernoulli utility $u$ over the prizes $\pi = (\pi_1, \ldots, \pi_K)$. Concavity of $u$ means that its slope is non-increasing, i.e.

$$\frac{u(\pi_k) - u(\pi_{k-1})}{\pi_k - \pi_{k-1}} \geq \frac{u(\pi_{k+1}) - u(\pi_k)}{\pi_{k+1} - \pi_{k-1}} \qquad \text{for } 1 < k < K \qquad (A.1)$$

Let $p$ and $q$ denote the probabilities defining lotteries $\mathscr{L}$ and $\mathscr{L}_k$ respectively. The difference $p - q$ has entries

$$(p-q)_\ell = \begin{cases} 0 & \text{if } \ell < k-1 \\ -\frac{\pi_{k+1} - \pi_k}{\pi_{k+1} - \pi_{k-1}} p_k & \text{if } \ell = k-1 \\ p_k & \text{if } \ell = k \\ -\frac{\pi_k - \pi_{k-1}}{\pi_{k+1} - \pi_{k-1}} p_k & \text{if } \ell = k+1 \\ 0 & \text{if } \ell > k+1 \end{cases}$$

If $\mathscr{L}$ is preferred to $\mathscr{L}_k$ for $1 < k < K$ then

$$u(\pi_k) \geq \frac{\pi_{k+1} - \pi_k}{\pi_{k+1} - \pi_{k-1}} u(\pi_{k-1}) + \frac{\pi_k - \pi_{k-1}}{\pi_{k+1} - \pi_{k-1}} u(\pi_{k+1}) \qquad \text{for } 1 < k < K \qquad (A.2)$$

and it is straightforward to verify that (A.2) is equivalent to (A.1). ∎

# B. Instructions

**WELCOME**

Thank you for participating in today's session.

The entire experiment will be conducted through your computer. Please avoid any distractions during the session.

---

**General instructions**

- During the experiment you can earn money.

- The amount earned may be different for each participant. The amount you earn depends on the choices made and on chance.

- In addition, you will receive a $10 show-up fee.

---

**General instructions**

- The experiment has two parts.

- At the end of the experiment, the computer will randomly select one choice from each part. You will be paid only those two choices.

- I will start explaining Part 1. When it is finished, please wait for further instructions.

---

**Part 1**

- You will choose between 2 different lotteries: A or B.

- You will choose between A or B ten different times.

- These ten choices are shown in ten different rows.

- The dollar amounts in lotteries A and B will always be the same.

- The probabilities in lotteries A and B will change from row to row.

---

**Part 1**

- For both lotteries A and B, the probability of the bigger amount starts at 10% and increases to 100%.

- For both lotteries A and B, the probability of the smaller amount starts at 90% and decreases to 0%.

- If you select B in a certain row, then you cannot select A in the rows below it.

---

**Part 1: payments**

- At the end of the experiment, the computer first randomly selects one of your choices (one of the rows).

- Then, the computer rolls a 100-sided die, which determines the payoff for the selected choice.

- For example, option A of row 1 pays $21 if the roll of the die is 1-10 and $16 if the roll is 11-100.

## Part 1

| | | | |
|---|---|---|---|
| 10% of $21.0, 90% of $16.0 | (A) | (B) | 10% of $38.5, 90% of $1.0 |
| 20% of $21.0, 80% of $16.0 | (A) | (B) | 20% of $38.5, 80% of $1.0 |
| 30% of $21.0, 70% of $16.0 | (A) | (B) | 30% of $38.5, 70% of $1.0 |
| 40% of $21.0, 60% of $16.0 | (A) | (B) | 40% of $38.5, 60% of $1.0 |
| 50% of $21.0, 50% of $16.0 | (A) | (B) | 50% of $38.5, 50% of $1.0 |
| 60% of $21.0, 40% of $16.0 | (A) | (B) | 60% of $38.5, 40% of $1.0 |
| 70% of $21.0, 30% of $16.0 | (A) | (B) | 70% of $38.5, 30% of $1.0 |
| 80% of $21.0, 20% of $16.0 | (A) | (B) | 80% of $38.5, 20% of $1.0 |
| 90% of $21.0, 10% of $16.0 | (A) | (B) | 90% of $38.5, 10% of $1.0 |
| 100% of $21.0, 0% of $16.0 | (A) | (B) | 100% of $38.5, 0% of $1.0 |

Please, choose between lottery A and lottery B in each case by clicking on the circular button A or B.
To confirm your choice, click on the OK button.

---

## Part 2

- This part consists of six periods.
- Each period has two choices, where you will choose between two different lotteries.
- The lotteries are represented in two ways.
- On the left you see the probabilities and the dollar amounts. On the right you see the same information represented by bars.
- The size of the bar represents the probability of obtaining the amount shown on top of the bar.

---

**Choice 1 and 2**

(A) 21% of $1.0 / 16% of $16.0 / 63% of $21.0

(B) 25% of $1.0 / 75% of $21.0

(C) 21% of $1.0 / 65% of $16.0 / 14% of $38.5

CHOICE 1: (C) or (A)

CHOICE 2: (A) or (B)

Please, choose by clicking the corresponding button. The bars are a graphical representation of the lotteries.
To confirm your choice, click on the OK button.

---

## Part 2: payments

- At the end of the experiment, the computer first randomly selects one of your choices in this part.
- Then, the computer rolls a 100-sided die, which determines the payoff for the selected choice.
- For the example on the previous slide, option A pays $1 if the roll of the die is 1-21, $16 if the roll is 22-37, and $21 if the roll is 38-100.